\documentclass[preprint2]{aastex}

\shorttitle{Progress on Studies of SNRs} \shortauthors{J. W. Xu}

\begin{document}

\title{Some Recent Progress on the Studies of Supernova Remnants}

\author{Jian-Wen Xu\altaffilmark{1}} \affil{Key Laboratory of Frontiers
in Theoretical Physics, Institute of Theoretical Physics, Chinese
Academy of Sciences, Beijing 100080, China} \email{xjw@itp.ac.cn}

\altaffiltext{1}{Postdoctor, Institute of Theoretical Physics,
Chinese Academy of Sciences, Beijing 100080, China.}

\begin{abstract}
We briefly reviewed some recent progress on the studies of supernova
remnants (SNRs), including the radio SNRs (the structure,
polarization, spectrum etc.), observational characteristics of X-ray
emission, pulsar wind nebulae (PWNe), association properties between
SNR and PSR, interaction of SNR and interstellar medium (ISM),
cosmos ray and the SNRs in external galaxies, etc.. Correspondingly
to the continue improvement of space and spectrum resolution of the
on-ground and in-space astronomical equipments at wavelengthes as
radio, optical, X-ray and so on, we know about SNRs more and deeper.
\end{abstract}


\section{Introduction}

SNRs are obviously the most bright radio source detected. The
firstly confirmed remnant reported by Bolton et al. (1949), i.e. the
Crab nebula connected with Taurus A. Thereafter more SNRs are
detected and studied at radio wavelength, and extensively analyzed
recently because of the observation at X-ray (review, Bernd \&
Aschenbach, 2002). SNRs have an important influences on the
properties of interstellar medium, to some great extent the
evolution of the host galaxy. They enhanced the abundant of heavy
elements in ISM. Their blast waves change and heat ISM, compress the
ISM magnetic field. Their shock waves effectively accelerate the
high energy cosmos rays, and so on. In Milk Way the confirmed SNRs
numbers have surpassed 270. Combining the observational radio data
and X-ray once can help to the establishment of the SNR model.

In general, shell-type supernova remnants evolve through four
stages: free-expansion phase, Sedov-phase, radiative-phase (or
snow-plough phase) and dispersion-phase. The SNRs age at first stage
is less than 200 years, their linear diameter is less than 1.3~pc. A
major of the detected SNRs are at adiabatic phase. SNRs at
first-phase or at fourth-phase are almost undetectable.

Researches to the remnants evolution in uniform ISM have already
been fruitful. SNRs evolution in stable stellar wind (with density
$\rho \propto r^{-2}$) has also been a studied topic. Chevalier \&
Liang (1989) analyzed the evolution in the bubble blown by
circumstellar wind. Numerical researches have been completed by
Tenorio-Tagle et al. (1991) and Franco et al. (1991). In the recent
few years, the numerical model about SNRs structure and evolution
have reach an unprecedented refinements. However, the analyzed model
is still play an important role when needing study the general
properties of supernova remnants, and needing to obtain the direct
connections between the purely observational parameters (for
example, the sizes, fluxes etc.) and the intrinsic physical
parameters.

Two useful SNRs catalogues could been found by internet. The first
one is the catalog of 275 SNRs edited through the literatures and
materials until the December, 2009 by David Green
(http://www.mrao.cam.ac.uk/surveys/snr), including the radio fluxes,
spectrum and conferences etc.. Another useful one of the SNRs
database store was provided by Sergei Trushkin
(http://cats.sao.com). It also includes the observational
information at optical, red-infrared and X-ray wavelengthes. Only
limited numbers of SNRs are detected at optical, red-infrared and
X-ray. We have detected more SNRs since the improvement of the
observational sensibility, in particular the ROSAT all-sky surveys
based on X-ray we detected a great deals of the SNRs candidates
(Busser, 1996). Therefore, the currently only 30\% per cent rate of
the SNRs at X-ray will increase following this, in spite of the
confinements induced by its short calculation time, relatively soft
energy spectrum and the remarkable absorbed efficiency to the
distant objects on Galactic plane. However, in many cases,
thereafter radio observations could confirm the supernova (SN)
origin of the SNRs candidates discovered by ROSAT survey (Schaudel
et al. 2002). The data in SNRs catalogs ether come from special
researches, or from the continue radio sky survey. How about the
completeness of the detected data? According to the statistics,
there should be 40 SN events in Milk Way in recent 2000 years, but
only 8 events had been observed. Tow predicted most highly energy
SNRs, i.e. the most bright radio sources are Cas A and Crab nebula,
born separately in C.D. 1680 and 1054. At present the sensibility at
radio sky survey is good enough to detect the SN burst sources under
the typical interstellar circumstances in the whole Galaxy. But
there has still been the problem that one can not distinguish the SN
origin SNR or the one not.

The history to study SNRs has surpassed half century. Woltjer (1972)
and Mills (1974) etc. had reviewed the research situation of the
Supernova remnants in detail. We briefly introduce some recent
progress in the studies of the SNRs based on the literatures after
the year 1990.

\section{Radio supernova remnants}

\subsection{Radio SNRs --- structure}

In general, the radio SNRs can be classified as 3 types: shell-type
(S-type), plerion-type (P-type) and composite-type (C-type). The
S-type remnants usually exist in uniform ambient medium, the
majority of the detected SNRs belong to this sort. C-type remnants
composed of a plerion component surrounding a shell outside. The
plerion remnants are irregular in form, and seldom be observed in
Galaxy.

SNRs are the nebulae origin from the massive stars outburst. The
released total energy is usually about $10^{53}$~ergs. Most of its
energy are carried away by neutrino. The neutrino origin at the
degeneracy debris (neutron star) forming stage after collapsing of a
protostar. Small part of the energy (about 1 per cent) lead to blast
waves. The waves sweep up the outer layer of the star, and producing
violent shock waves traveling through the ambient medium. The
details of the expanding process depend on a group of different
physical parameters --- the total mass and energy of the ejections,
the distribution density and property in surrounding medium
(Dwarkadas \& Chevalier 1998; Featherstone et al. 2001; Blondin et
al. 1996), and the energy loss rate of decreases of the neutron star
rotation and the velocity of proper motion (Chatterjee \& Cordes
2002; Frail et al. 1994). Therefore in principe we can foresee the
structure variety in the pulsar wind nebulae (PWN)-SNR system.

In general, the supernova remnants take on a spherical form, but in
fact there is always a deviation from that form. The deviation
extent will be bigger when the SNR shock waves pushing upon a dense
nebula (Levenson, Granham \& Walters 2003). Larger deviation means
bigger density of the ISM. The shock waves formed in the stellar
nebula generally emit radiations, inducing the cooling of the shock
waves gas. Most of the energies are vanished by the radiative
transformation. The large deviation of the spherical form could be
caused by the gas blow-out when the shock waves collide with a block
of dense molecular clouds.

Some SNRs show fill-center structure, a typical example is the Crab
nebula, but we also currently discover other similar SNRs. They are
called Plerions, or Crab-like SNRs. They are so called Plerionnot
only because of their forms, but also their other properties, i.e.
flat power law spectrum at radio wavelength, the spectra index
ranging from 0.0 to 0.3; highly radio polarization, owning neatly
form, but not all the plerions show such; the power law spectrum at
X-ray, the energy spectrum index is near about $-$2 (Asaoka \&
Koyama 1990); A detected PSR associated with it (some plerions have
no PSR).

Although we do not know much in very details about the properties
and structure of the plerions, but the followings are well-known:
plerion is a bubble in expanding, mainly composed of the magnetic
field and relativistic electrons, the detected synchrotron
radiations come from this two components; in order to explain the
typical synchrotron emissions and the higher frequency radiations,
one introduces continue magnetic fluxes and relativistic electrons.
At the high frequency radiations, the life time of the synchrotron
particles is obviously shorter than the SNR evolution ages.

Regarding a simple approximation, we suppose that in the plerions
bubble the magnetic field and particles distribution is uniform: it
is usually enough to explain the whole spectrum evolution of the
nebula. However, the uniform suppose seems to be not correct. The
newly born particles released from the associated pulsar, justly
located near the terminal shock waves of the PWN. Therefore the
uniform degree relies on the validity of the propagating particles
among the remnant.

Furthermore, the magnetic field structure may be very complicated.
Regarding from the magnetohydrodynamics (MHD), the spherical model
is not enough to interpret the magnetic field structure: it probably
have the cylindrical symmetry (Begelman \& Li 1992), but more
complex structure type is detected. Important information of the
plerion structure can be gotten by comparison of the high resolution
images at different wavelengthes, and also provide the clue which
ruled the process of the magnetic field evolution and particles
distribution.

\subsection{Radio SNRs --- polarization}

Usually SNRs own linear polarization about 10\%-20\%. Therefore the
linear polarization exits or not is another useful criterion about
weather a source is a SNR or not.

Soon after detecting the linear polarization from the radio object
sources, we easily discover that the whole galactic disc is an
important source of the polarization emissions (Seeger \& Westerhout
1961). This sort of radiations have at least two components (Duncan
et al. 1997a) --- a polarized emission coming from discrete
remnants, and another more discrete background polarization
radiation. The later produced from the interactions between the
relativistic electrons in ISM and the Galactic magnetic field.

The electron synchrotron emissions in the uniform magnetic field own
highly orientation. Therefore they should be linear polarization
(Moffett \& Reynolds 1994a). In the ideal case, the magnetic field
location can be determined by the polarized position angle. Here we
assumed the vacuum magnetic field is a regular one. We now have
already known that the comparatively younger SNRs owning relatively
larger depolarization affection, which is caused by the irregular
arrangement of the magnetic field of the source itself (Moffett \&
Reynolds 1994a, b).

The polarization degree is an important physical parameter on the
supernova remnants, because it is an indicator about the magnetic
field regularity. For uniformly distributed magnetic field, the
synchrotron polarization degree (P) directly connected with the
spectrum index ($\alpha$) and do not relied on its frequency: $P =
(3 - 3\alpha)/(5 - 3\alpha)$.

\subsection{Radio SNRs --- spectrum}

In Galaxy the SNRs calculated radio continue spectrum from meter
wavelength to centimeter wavelength or more shorter wavelength
usually takes on a power law. The spectrum index scope is $\alpha
\approx - 0.5$ ($S \propto \nu^{\alpha}$) for shell-type remnants,
and $\alpha \approx - 0.1$, and also many mixed radio radiations
with middle spectrum value. However, at the frequency lower than
100~MHz, Lacey et al. (2001) point out that about two third of the
SNRs show broken-spectrum, denotes the thermal absorption. The
continuum optical depth and spectrum broken has no connection with
the very uncertain remnants distance. All these above do not support
the point of view that the absorption comes from the spherically
distributed thermally ionized medium (with number density $n \sim
0.1$~cm$^{-3}$). The instead view is that the absorption must come
from the local ionized zones with comparatively larger density ($n
\geq 1$~cm$^{-3}$), but the zones size is still unknowable, the
filled factor very small ($\leq 1\%$). Therefore the reasonable
interpretation should be that the absorption material is the
extended HII areas (EHEs), the ionized gas surrounded by a normal
HII region, just as the result derived through observational
comparison between the centimeter waves and meter waves radio
recombined lines (RRL). The absorption also may be induced by the
superimposed materials of many small normal HII areas or planetary
nebulae.

Among the SNRs basic physical parameters --- the distance to the
observer, linear diameter, Height to the Galactic disc, evolved age,
luminosity, fluxes density and radio spectrum index, etc. the most
important and significant one is the spectrum index. The remnants
radio spectrum takes on the power law which indicate that the SNRs
radiation process is not the thermal emissions but the synchrotron
one. This is very important in the SNRs physics.

\section{X-ray radiations}

Supernova remnant is a protagonist among the interstellar materials.
They eject high energy substances and emit cosmos rays. Because the
SNR shock waves velocity reaches the scope of hundreds and thousands
of kilometers per minute, the gas is heated to the high temperature
of some millions degree. Therefore the majority of emissions is at
X-ray. X-ray observations become the best method to study the gas
mass.

In 1999 new generation of X-ray telescopes (Chandra \& XMM-Newton)
had been launched, providing good space and spectral resolution for
the first time, which make it possible to plot the emission line
spectrum of the heavy mental elements. Among the young SNRs, this
could give a clue to the nuclear synthesize of the SNR protostar,
make a theoretical estimates of their main elements absolute
contents and the distribution of their radius and azimuth angles.

With its angle resolution of arc-second, Chandra could make clear
about the radial movement of gas shock waves of the young remnants.
The radial distribution profile of the matter densities and
temperature at the interval space of the traveling shock waves and
reflection shock waves, is directly corrected with the distribution
of matter densities and temperature after the SNe outburst.

The foresee products of a supernova outburst from the nucleus
collapse of a massive star includes a shock waves expanding to the
interstellar or circum-stellar medium, expanding post-shockwave
mental-rich ejector from the protostar, and relic from the star
nucleus collapse (usually a rapid-rotated magnetic neutron star).
These relics show a bright shell with radio emissions (being
out-moving shock waves) and the hot shockwave gas inside the shell,
protostar ejector emitting thermal X-ray, a short-period
radio/X-ray/$\gamma$-ray pulsar and a central bright nebula, which
formed by the relativistic out-flow particles of a pulsar and
pushing outside. Until now more than 1300 pulsars in Galaxy have
been detected (Manchester et al. 2002), more than 230 radio SNRs
(among them more than 80 SNRs emitting X-rays), and about 24 pulsar
wind nebulae (PWNe, Kaspi \& Helfand 2002).

Recently the observational images higher than 3.5~keV leads to the
discovery of the X-ray synchrotron emission nebulae in the remnant.

The X-ray observations of the Crab-like and composite-type SNRs
provide important information of the less-known original
distribution of the pulsar magnetic field, rotational period and
ages, supernova remnants dynamics etc..

The Crab Nebula plays an important role for us to know the pulsar
wind nebula. The basic X-ray observational characteristics is a
pulsar at center, an ellipse ring and two out-flow ejectors but
without shell around. The X-ray ring is the post-shockwave
equatorial wind coming from the central pulsar (Weisskopf et al.
2000), and the out-flowers come from the pulsar two rotational polar
regions (Aschenbach 1992). Observational characters and its models
let us know more about the high energy and geometry properties of
the PWNe and the physics of extra-relativistic shock waves and
particle acceleration.

However, the Crab morphologic is rather unique. We have not
discovered any other SNRs owning all of the X-ray basic
observational characters like the Crab nebula until now (Gaensler
2001). The PWNe around Vela pulsar and PSRB1509-58 have X-ray arcs,
but these arcs are not the complete once like the Crab, and both
them located in the diffuse and bright hot ejector (Helfand et al.
2001; Gaensler et al. 2002).

Through the X-ray spectrum analysis we know that the central
component emissions of a majority of the composite-type SNRs is the
thermal radiation. Therefore White \& Long (1991) figured out a
physical model showing that the SNRs character is the X-ray thermal
emission with a peaked center, which caused by the vaporized cloud
after the remnants blast waves travel through. This model had been
applied to some SNRs, although debates still be there about it is
true or not.

More and more Galactic SNRs revealed the non-thermal emissions at
X-ray. The non-thermal emission do not connect with a pulsar. These
emissions could occupied the most or entire remnants, such as
follow: SN1006 (Dyer et al. 2001), G266.2$-$1.2 (RXJ0852.0$-$4622;
Slane et al. 2001), G347.3$-$0.5 (Slane et al. 1999; Uchiyama et al.
2003) and AXJ1843.8$-$0352 (Ueno et al. 2003). In the sensitive
range of RXTE PCA hard X-rays (until 60~keV), the thermal radiations
of some SNRs (Cas A, Kepler, Tycho, SN1006 and RCW86) were detected.
A majority of the cases among these examples confirmed that the
X-ray non-thermal emissions are synchrotron radiations.

SNRs are the sources which the high cosmos rays most likely come
from. Its evidence is that we detected the X-ray synchrotron
emissions from some SNRs shell (Koyama eta l. 1995, 1997; Slane et
al. 2001), and detected the TeV energy rank $\gamma$-rays from some
SNRs (Tanimori et al. 1998; Muraishi et al. 2000; Enomoto et al.
2002). Electrons are accelerated to high energy about 1~TeV or
higher, the acceleration mechanism possibly is one order
Fermi-acceleration. Because the energy enhanced rate of the
accelerating electrons is proportion to the magnetic field $B$, but
the energy loss rate of synchrotron radiations is proportion to
$B^2$, the high energy electrons producing the X-ray synchrotron
emissions most likely exit in the SNRs shell with weak magnetic
field, where the radio fluxes are very low (radio fluxes are
proportion to $B^2$).

\section{Pulsar wind nebulae (PWNe)}

When the pulsar relativistic wind of a pulsar is constrained, a
synchrotron pulsar wind nebula (PWN) is formed. Because the life
time of the synchrotron high energy electrons are very short, we
could directly trace the pulsar energy fluxes from the X-ray
emissions of a PWN. Therefore, from the spectrum and morphological
characters of a X-ray PWN we could reveal the PWN structure and
chemistry buildup, even extrapolate the direction of the pulsar
self-rotated axis or the velocity vector. For the PWN which the
associated pulsar is not detected, only through observations to the
PWN radiations can we peek into the position and energy of the
source interior.

There was such a PWN just at the center of the SNR G0.9+0.1 without
detected pulsar inside. This PWN firstly confirmed at radio
wavelength, but recently it is also detected at X-ray by $BeppoSAX$
(Mereghetti et al. 1998; Sidoli et al. 2000). On these observations
the X-ray emissions show the power-law spectra with its energy
spectral index of $\Gamma = 2.0\pm 0.3$ (Here $N\propto
E^{-\Gamma}$), the fluxes at $2- 10$~keV of $f_x =
6.6_{-0.8}^{+1.3}\times 10^{-12}$~ergs~s$^{-1}$~cm$^{-2}$.

The most well-known PWN driving by the pulsar is the Crab nebula.
Observations by long terms show that the Crab is very different with
a majority of other SNRs. It is compounds of a synchrotron nebula
and an associated pulsar, but without any shell around outside.

\section{Association of SNRs and PSRs}

We all know that both the SNRs and PSRs are formed from the
supernova outburst. However, the events both them associated are
very seldom. Such events could provide very important information of
both connection and interaction. In particular, for the case that
the associated pulsar is still inside the remnant, the high pressure
could constrain the pulsar relativistic wind. This leads to the
synchrotron emission, and a detectable PWN is formed.

In nearly 15 years, we know that only two SNRs, the Crab Nebula and
the Vela remnant, are associated with the pulsar. The third SNR-PSR
associated example only was found out after the detection of a
X-ray, $\gamma$-ray and radio pulsar, PSR B1509$-$58, with
150~millisecond period inside SNR G320.4$-$1.2 (MSH 15$-$52). The
PSR rotation parameters were used to estimate the characteristic age
of the associated SNR of $\tau_c = 1700$~yr.

In the past years, the SNRs and PSRs associated examples have
enhanced to near 30, as the result of following series of new
detected methods ---- search survey for the young PSRs at high
frequency, the mutual correlation of SNR catalogue and PSR atlas,
intensively search for the PSRs or SNRs around the SNRs or PSRs
(Lorimer et al. 1999).

Association of SNRs and PSRs make it possible to measure some
physical parameters of both. But it is not so when both them
separately exit. Because we assume the PSR born at the remnant
center, combining the PSR characteristic age and the deviation of
the PSR from the SNR, one could estimate the transverse velocity of
the pulsar (Frail et al. 1994). Measure the direction of the proper
motion of this pulsar could confirm or exclude the pulsar associated
with SNR or not, the proper motion extent could independently fix
the PSR age (Gaensler et al. 2000).

\section{Interaction between SNR and ISM}

The interactions between the remnants shock waves and the stellar
clouds is a basic topics on interstellar gas dynamics which plays an
important role in studying the ISM evolution. The main physical
problems include following some: 1) How much are there the mass rate
and total mass of the nebula swept-up by shock waves? 2) How much
are there the moments transformed to the nebula? 3) How about the
outlooks of the nebula disturbed by shock waves? How about its
morphological and velocity distribution? 4) What role is played to
the nebula evolution by the element of vortex dynamics? 5) Could the
interactions with ISM lead to the new generation stars formation?
Highly none-linear interaction studies could not solve these
problems, numerical methods are needed to multi-dimensionally and in
details to study the shock wave nebula dynamics. Klein et al. (1994)
make use of the later method discovering that the gas clouds could
been destroyed by series of gas instabilities disturbed by the shock
waves. Recently, Levenson et al. (1997) observed the Cygnus Loop
remnant in details making use of the ROSAT high resolution imager
(HRI). In Cygnus Loop the interactions between the ISM and shock
waves play an important role, and providing important observational
evidences for the key role played on the remnants morphologic by the
ISM non-uniformity.

Supernova remnants are regarded as the energy sources of the local
ISM dynamics, forcing the gas constantly move on and returning the
dense clouds material to the more dispersion interstellar medium or
galactic halo. The violent shock waves traveling among the
interstellar clouds compress and heat interstellar materials, change
their chemistry components and induce the star formation. Although
we foresee the tightly association of the II type SNe and clouds, we
currently know only a few examples of the supernovae and clouds
interaction. At the molecular clouds edge or nearby, the blast waves
would quickly pass through the thin regions of the mass medium and
push the slowed shock waves into the denser medium mass. For the
SNRs inside dense matter, the infrared rays are in particular useful
diagnose tool, because the infrared rays fine structure were born
from the dense gas behind the shock waves with radiations. It also
could track and measure all the abundances of the ground-state
elements.

Magnetic field can buffer the action of the shock waves to the
clouds. In the case without magnetic field, the SN blast waves could
heat, compress and decompound or even totally destroy the clouds at
last (Klein et al. 1994). But the intervening magnetic field could
deduce this effects, limit the compression, enhance the stability
and the gas cloud remains there, and even induce the formation of a
new generation stars (Miesch \& Zweibel 1994; Mac Low et al. 1994).

Making use of the OH molecular emission line at 1720.53~MHz, Brogan
et al. (2000) explored the interactions between SNR and molecular
clouds. We have already detected the OH maser of about 20 SNRs which
occupy about 10\% of the total known SNRs (Koralesky et al. 1998).
This sort of maser lines are produced through the collision with
$H_2$ ($n\sim 10^4$~cm$^{-3}$, $T\sim 80$~K) behind the shock waves
of composite-type SNR. Only by the side of the shock wave fronts one
can detect the strong maser lines, this constrain our judgement to
the shock waves geometry. These describes on observation support the
OH-(1720~MHz)-excite-theory-model (Lockett et al. 1999; Wardle et
al. 1999; Wardle 1999).

The Crab Nebula, 3C58 and other plerions are without any rapidly
ejecting shell or conora. This is possibly because that they located
inside a low density ambient medium and can not form the detectable
shock waves. The HI bubble surrounding some plerions are the
evidence supporting this assumption (Wallace et al. 1994). But the
outcome of SNR G21.5$-$0.9 is still shown uncertain.

\section{SNRs and cosmos rays}

SNRs are regarded as the first candidates of the cosmos ray
accelerators. They are able to enhance the cosmos rays to the energy
valve point ($E \approx 3000$~TeV). Although the relation between
the SNRs and cosmos rays have been extensively concerned (Jones \&
Ellison 1991; Malkov \& Drury 2001), a majority of them are limited
on the theoretical studies, only in minor cases the observational
data are considered. One of the important unsolved issues concerning
with the SNRs accelerating cosmos rays is that how much the highest
energy values $E_{cutoff}$ can the cosmos rays particles be
accelerated by the remnant shock waves fronts. For the young
shell-type SNRs in Milk Way and Large Magellanic Cloud (LMC) the
highest energy values of the accelerated cosmos rays electrons
$E_{cutoff} \leq 200$~TeV (Reynolds \& Keohane 1999) and $E_{cutoff}
\leq 80$~TeV (Hendrick \& Reynolds 2001), both them are greatly
lower than the valve point of the cosmos rays energy spectra. The
angle resolution and spectrum resolution shortages of the remnants
X-ray observations limited the progress on the acceleration
processes researches.

However, since the launches recently of the X-ray astronomical
satellites such as ROSAT, ASCA, Chandra and XMM etc., there are very
high resolutions on the SNRs observations at X-ray energy scope.
These high resolution observations are useful for the high space
resolution spectrum studies of the X-ray emissions from SNRs,
especially for the investigates and studies on the edge areas of
bright luminosity emissions corrected with the SNRs expanding shock
wave fronts. At the shock wave fronts the interstellar or
circum-stellar media are swept up, where one could explore the
cosmos rays acceleration process by the details spectrum analysis to
the X-ray emissions from the SNRs edge regions. Most original works
make use of the ROSAT, ASCA and RXTE observations to study the high
energy X-ray radiation from Cas A and SN1006 (Willingale et al.
1996; Allen et al. 1997; Keohane 1998; Dyer et al. 2001; Allen \&
Gotthelf 2001). For more, to study the SNRs at more complete
wavelengthes is needed in order to understand the cosmos rays
acceleration process in details.

In fact, only recently the $\gamma$-ray-image-telescope based on the
ground directly provides the evidence that the TeV high energy
relativistic particles (most probably the electrons) exit among the
shock waves of 3 shell-type SNRs: SN1006 (Tanimori et al. 1998), Cas
A (Aharonian et al. 2001) and RXJ1713.7$-$3946 (namely G347.3$-$0.5;
Muraishi et al 2000). Although the former reports (Aharonian \&
Atoyan 1999; Combi et al. 2001) ponder that the detected
$\gamma$-rays at the SNR direction are possibly because of the
baryon interactions, but this can not exclude that the detected
radiations come from the high energy electrons or the nearby pulsar
(e.g. Brazier et al. 1996; De Jager \& Mastichiadis 1997; Gaisser et
al. 1998).

\section{SNRs in external galaxies}

In general, we emphasize on looking for the SNRs in Galaxy. But
because the absorption and extinction of the Galactic disk
interstellar media and the Sun location at the galactic plane these
sorts of researches are very difficult. On the contrary, it would be
better to look for the SNRs in external galaxies, especially the
near face-on and high galactic latitude galaxies, its interior
absorption and Galactic absorption are very less.

The already known SNRs in M31 and M33 are more than 300. Such great
deal of database provide quite important clue for us to solve the
key problem of the SNRs evolution and the interaction between SNR
and ISM in M31 and M33.

In recent more than 20 years, we have already done extensive
researches to the SNRs X-ray radiations from the Galaxy. For many
SNRs the observational space resolution before the Chandra launch is
fine enough which had discovered many different sorts of the SNRs
morphologic. However, the Galactic SNRs researches are constrained
because the lack of the reliable estimated distance values and
usually the great absorption of the interstellar medium. There would
be no such limit to study the external galaxies SNRs. Chandra can
decern the SNR structure in details in the nearest galaxies (LMC and
SMC) (for example, Hughes 2001).

M31 is the nearest galaxy with Galaxy-like shape and size ($\sim
800$~kpc). It is the best candidate to study the SNRs by comparison.
The typical method for us to confirm SNR in M31 through optical
survey is based on the [SII] and $H_{\alpha}]$ observational images
(e.g. Braun \& Walterbos 1993; Magnier et al. 1995). These surveys
have already detected about 200 SNR candidates, among them 27 SNRs
have been confirmed by the spectroscope measures. Recently, Supper
et al. (2001) confirmed 16 SNRs on the area of $\sim 10.7$ square
degree by the extensive ROSAT Position-Sensitive-Proportion-Counter
(PSPC) survey to M31 and by the correlation with optical atlas.
Their X-ray luminositys (0.1$-$2.4~KeV) range from $\sim 10^{36}$ to
$\sim 10^{37}$~erg~s$^{-1}$.

More recently, Kong et al. (2002) detected 2 SNRs (XOCM31
J004327.7+411829 and CXOM31 J004253.5+412553) in the region of $\sim
17'\times 17'$ at the M31 center with Chandra. Both SNRs had once
been confirmed before by ROSAT. It is worthy to mention that XOCM31
J004327.7+411829 had also once been detected by Einstein
Observatory.

One of the aims to study SNR is to explore the protostar properties
by making use of the SNR related characters. This characters, for
example, the rich Fe abundance in the partial neutral gas or
ejectors at the ambient medium have been taken to illustrate that
some SNRs origin from Ia Type supernova. Hughes et al. investigate
the SNR DEML71 (0505-67.9) in the LMC. This remnant owns the two
characters just mentioned above. Based on the Einstein Observatory
X-ray survey on LMC, Davies et al. (1976) pointed out at first time
that DEML71 is a supernova remnant. Thereafter, the optical
spectroscope confirmed it for further more, which shows the SNR
emission lines are mainly the Balmer lines (Smith et al. 1991). In
other words, the optical spectrum of the SNR is mainly the hydrogen
emission lines, but seldom or without [0III] or [SII] forbid lines
radiations.

\section{Conclusion}

The high resolution images comparison among different frequencies
could provide rich information about the plerions, helping to make
clear some publicly well-known issues on this sort of objects. For
example, how, where and how many particles with different radiation
processes in the plerion interior are accelerated? How do they
propagate in the nebulae? How about the magnetic field structure?
How do the evolution of the plerions and their associated neutron
star affect the synchrotron emission?

The great step toward to help understanding these topics is the
emerge of the new generation of X-ray telescopes. They have the
angle resolution of arc-second, forming spectrum images. The life
time of the X-ray electrons is quite short, typically the time which
the light travels through a nebula. Therefore, we could directly
gain the position information where currently the particles pushing
into from the X-ray spectrum image.

The millimeter wavelength is another wavelength scope which could
effectively detect the remnants physics. One of the different
spectrum broken point just located near this wavelength. Therefore,
the spectrum figures could indicate the varied situation of the
spectrum space at broken point, and offer the structure information
of the magnetic field.

For composite-type remnants and plerions the high frequency and high
resolution polarization observation is in particular important. But
despite of the great efforts, the detected source numbers are still
very few. In order to constrain some parameters of the weaker
sources among the two sorts of SNRs, high sensitivity observations
are demanded.

Despite of the great numbers of currently known shell-type SNRs,
there are still more and more SNRs sources continuously being
confirmed, especially the large diameter and low surface brightness
objects. Confirming these sources at high frequency is ether
difficult or consuming lots of time. At the longer wavelength, the
combined data of the aperture synthesis telescopes and the single
antenna could reduce the constraint by the mixing background sources
from external galaxies, which is a successful method.


\end{document}